\documentclass[12pt]{article}

\usepackage{amsfonts,amsmath,amssymb}
\usepackage{hyperref}
\usepackage{cite}
\usepackage{epsfig}
\usepackage{paralist}
\usepackage{fancyhdr}
\usepackage{tikz}
\usepackage{tkz-euclide}
\usetikzlibrary{decorations.pathmorphing,patterns,calc,snakes,arrows}

\usepackage{graphicx}
\usepackage{xcolor}
\numberwithin{equation}{section}
\usepackage[vcentermath]{youngtab}
\usepackage{etex}
\usepackage{braket}
\usepackage{float}

\def\spa#1{\phantom{\fbox{\rule[-#1cm]{0cm}{0cm}}}}

\def\be{\begin{equation}}
\def\ee{\end{equation}}
\def\bea{\begin{eqnarray}}
\def\eea{\end{eqnarray}}

\def\Tr{\mbox{Tr}}
\def\del{\partial}

\renewcommand{\thefootnote}{\fnsymbol{footnote}}

\makeatletter
\g@addto@macro\bfseries{\boldmath}
\makeatother

\def\p{\partial}

\def\zb{{\bar{z}}}

\def\Tb{{\bar{T}}}

\def\Tr{\mathop{\mathrm{Tr}}\nolimits}

\begin{document}

\hfuzz=100pt
\title{{\Large \bf{$T\Tb$ braneworld holography}}}
\date{}
\author{Shinji Hirano$^{a, b}$\footnote{
	e-mail:
	\href{mailto:shinji.hirano@gmail.com}{shinji.hirano@gmail.com}}
  \,and Vinayak Raj$^{a}$\footnote{
	e-mail:
	\href{mailto:vinayak.hep.th@gmail.com}{vinayak.hep.th@gmail.com}}}
\date{}

\maketitle

\thispagestyle{fancy}
\rhead{YITP-25-122}
\cfoot{}
\renewcommand{\headrulewidth}{0.0pt}

\vspace*{-1cm}
\begin{center}
$^{a}${{\it School of Science, Huzhou University}}
\\ {{\it Huzhou 313000, Zhejiang, China}}
  \spa{0.5} \\
$^b${{\it Center for Gravitational Physics and Quantum Information (CGPQI)}}
\\ {{\it  Yukawa Institute for Theoretical Physics, Kyoto University}}
\\ {{\it Kitashirakawa-Oiwakecho, Sakyo-ku, Kyoto 606-8502, Japan}}
\spa{0.5}  

\end{center}

\begin{abstract}
We study a constructive gravitational dual of two-dimensional $T\Tb$-deformed conformal field theories (CFTs) grounded in their two-dimensional gravity description.
This framework can be viewed as a Randall-Sundrum-type braneworld, where two-dimensional gravity localized on the $AdS_3$ boundary is dynamical.
Assuming the AdS/CFT correspondence, the holographic dictionary provides a straightforward translation between the $T\Tb$-deformed CFT and its gravitational dual. 
In particular, this $T\Tb$ braneworld holography remains valid in the presence of matter.
To clarify how our framework relates to -- and differs from -- the cutoff AdS proposal, we examine, in the absence of matter, the effect of dynamical boundary gravity on the bulk geometry. In general, semiclassical integration of the boundary two-dimensional gravity -- weighted by the massive gravity action -- leads to a deformation of the two-dimensional metric on constant-radial surfaces. However, we find that the surface commonly interpreted as the cutoff emerges dynamically as a characteristic bulk surface on which the deformation is neutralized.
This perspective opens the door to possible extensions of our framework to incorporate matter effects on the bulk surface in future work.
\end{abstract}

\renewcommand{\thefootnote}{\arabic{footnote}}
\setcounter{footnote}{0}

\newpage

\tableofcontents


\section{Introduction}
\label{Sec:Introduction}

The $T\bar{T}$ deformation~\cite{Zamolodchikov:2004ce,Smirnov:2016lqw,Cavaglia:2016oda} provides an instructive case of an irrelevant deformation of quantum field theory (QFT) that remains well-defined and exactly solvable.\footnote{A recent comprehensive review can be found in~\cite{He:2025ppz}.} One of the most illuminating perspectives on this deformation is its interpretation as a coupling to two-dimensional gravity~\cite{Dubovsky:2017cnj,Dubovsky:2018bmo,Tolley:2019nmm}. Within this framework, $T\bar{T}$-deformed QFTs can be reformulated as quantum field theories on fluctuating geometries governed by topological or massive gravity in two dimensions.

In the holographic setting, the $T\bar{T}$ deformation of conformal field theories (CFTs) has been conjectured to correspond to the introduction of a finite radial cutoff in the AdS bulk~\cite{McGough:2016lol}. This cutoff AdS proposal serves as an effective description in the pure gravity sector, successfully reproducing the energy spectrum and stress tensor data. However, the interpretation is incomplete: it fails to consistently incorporate bulk matter fields. In particular, when primary operators in the boundary theory -- corresponding to matter fields in the bulk -- are included, the naive cutoff prescription does not yield the correct boundary correlators. Moreover, it misses essential features of short-distance physics below the $T\bar{T}$ scale~\cite{HR_NP_Planck}.

In our previous work~\cite{Hirano:2020nwq,Hirano:2024eab}, partial progress was made in addressing these limitations by exploring the gravitational interpretation of the $T\bar{T}$ deformation using random geometry and dynamical coordinates. However, a full understanding of how matter fields are incorporated in the holographic dictionary has remained elusive.

In this paper, we propose a constructive gravitational dual of $T\bar{T}$-deformed CFTs grounded in their two-dimensional gravity description. This framework can be naturally viewed as a Randall-Sundrum-type braneworld~\cite{Randall:1999vf, Karch:2000ct}, where two-dimensional gravity is dynamical on the $AdS_3$ (UV cutoff) boundary~\cite{Gubser:1999vj}. Assuming the standard AdS/CFT correspondence~\cite{Maldacena:1997re,Witten:1998qj,Gubser:1998bc}, we demonstrate how the holographic dictionary continues to hold even when matter is included, extending our random geometry approach~\cite{Hirano:2020nwq} to finite $T\bar{T}$ coupling and  offering a more complete realization of $T\bar{T}$ holography.

To clarify how this construction relates to -- and differs from --  the widely assumed cutoff AdS proposal, we analyze the semiclassical integration over the boundary geometry in the absence of matter. We find that the surface commonly interpreted as the radial cutoff emerges dynamically as a characteristic bulk surface on which the $T\bar{T}$ deformation is neutralized. This dynamical realization sharpens the gravitational interpretation and paves the way for incorporating matter consistently within the holographic dual.

The paper is organized as follows. Section \ref{sec:FT} reviews the interpretation of the $T\bar{T}$ deformation as a coupling to two-dimensional massive gravity and summarizes supporting evidence. Section \ref{sec:GD} presents a constructive gravitational dual via Randall-Sundrum-type braneworld holography, where two-dimensional gravity is dynamical on the $AdS_3$ boundary, and introduces the holographic dictionary and stress tensor in the presence of bulk matter. Section \ref{sec:cutoff} examines the connection to the cutoff AdS proposal by analyzing the role of dynamical boundary gravity in the absence of matter, showing that the radial cutoff surface emerges dynamically as a characteristic bulk surface where the $T\bar{T}$ deformation is effectively neutralized -- verified to all orders in the coupling via a saddle-point analysis. Section \ref{sec:Discussion} summarizes our constructive $T\Tb$ braneworld holography. It also outlines promising avenues for future research. Appendix A details the derivations of the saddle-point solutions discussed in Section~\ref{sec:cutoff}.

\section{Massive gravity description of $T\bar{T}$-deformed CFT: A quick review}
\label{sec:FT}

To set the stage for our constructive approach to a gravitational dual of $T\bar{T}$-deformed CFTs, we begin by reviewing a formulation of the $T\bar{T}$-deformed theory that offers a particularly illuminating perspective for our purposes. Specifically, we adopt the interpretation of the $T\bar{T}$ deformation as a coupling to two-dimensional gravity~\cite{Dubovsky:2017cnj,Dubovsky:2018bmo,Tolley:2019nmm}, and work within the massive gravity framework developed by Tolley~\cite{Tolley:2019nmm}.
This framework is defined by the action\footnote{We adopt the convention $\epsilon^{01} = \epsilon_{01} = 1$ for the Levi-Civita symbol.}
\begin{align}\label{FT_action}
S_{T\Tb}[e, f]
=S_{\rm CFT}[e]+{1\over 2\mu}\int d^2x\epsilon^{ij}\epsilon_{ab}\left(e^a_{i}-f^a_{i}\right)\left(e^b_{j}-f^b_{j}\right)
\equiv S_{\rm CFT}[e]+S_{\rm mG}[e, f]\ .
\end{align}
Here, $e$ and $f$ are zweibeins corresponding to the two-dimensional metrics
$ds_{\rm CFT}^2 = h_{ij} dx^i dx^j$ and
$ds_{T\bar{T}}^2 = \gamma_{ij} dx^i dx^j$, respectively. 
Importantly, the metric $h_{ij}$ (or equivalently, the zweibein $e^a_i$) is dynamical, rendering the theory a form of two-dimensional quantum gravity.
The presence of the two metrics reflects two complementary descriptions of the
$T\bar{T}$-deformed theory: either as a deformed CFT living on a fixed
(undeformed) background with metric $\gamma_{ij}$, or as an undeformed CFT
living on a dynamically fluctuating geometry with metric $h_{ij}$. This perspective was proposed in~\cite{Conti:2018tca, Cardy:2019qao} and further developed, for example, in~\cite{Hirano:2024eab}.

The observables of the $T\bar{T}$-deformed CFT are related to those of the undeformed CFT through the generating functional of the deformed theory, given by
\begin{align}
Z_{T\Tb}[f, J]=\int{\cal D}e^a_i e^{-S_{\rm mG}[e, f]}Z_{\rm CFT}[e, J]\ ,
\end{align}
where $J$ collectively denotes the sources for the operators in the theory. This functional solves the diffusion equation or ``Schr\"odinger equation,''
\begin{align}\label{qFlow}
{\del\over \del\mu}Z_{T\Tb}[f,J]=\int d^2x \lim_{y\to x}{1\over 2}\epsilon_{ij}\epsilon^{ab}{\delta^2Z_{T\Tb}[f, J]\over \delta f^a_{i}(y)\delta f^b_{j}(x)}
\end{align}
which arises from the random geometry interpretation of the $T\bar{T}$ deformation at infinitesimal coupling~\cite{Cardy:2018sdv}.

As further evidence for the equivalence between $T\bar{T}$-deformed CFTs and the massive gravity formulation, we examine the properties of the stress tensor in the deformed theory. In our convention, the stress tensor of the $T\bar{T}$-deformed theory is defined as
\begin{align}
T_{ij}={2\over\sqrt{\det\gamma}}{\delta S[\varphi, \gamma_{ij}]\over \delta \gamma^{ij}}\qquad\mbox{with}\qquad \gamma_{ij}=\delta_{ab}f^a_{i}f^b_{j}\ .
\end{align}
In this setup, it is convenient to work with the stress tensor carrying mixed indices -- one associated with curved spacetime and the other with the local Lorentz frame -- given by
\begin{align}
\label{FT_d_stress}
T^{i}_a=f_a^{j}T_{j}^{i}= {1\over \det f^a_{i}}{\delta S_{T\Tb}[e, f]\over \delta f^a_{i}}
= -{1\over \mu\det f^a_{i}}\epsilon^{ij}\epsilon_{ab}\left(e^b_{j}-f^b_{j}\right)\ .
\end{align}
A key consistency check is to understand how the $T\bar{T}$ flow equation for the stress tensor is encoded in this framework.
First, observe that the definition of the stress tensor implies
\begin{align}\label{FT_detT}
2\det T^{i}_a={1\over\mu^2(\det f^a_{i})^2}\epsilon^{ij}\epsilon_{ab}\left(e^b_{j}-f^b_{j}\right)\left(e^a_{i}-f^a_{i}\right)\ .
\end{align}
It then follows that the trace of the deformed stress tensor can be expressed as
\begin{equation}
\begin{aligned}
T^{i}_{i}&=f^a_{i}T^{i}_a
=2\mu\det T^{i}_{j}
-{1\over \mu\det f^a_{i}}\epsilon^{ij}\epsilon_{ab}\left(e^b_{j}-f^b_{j}\right)e^a_{i}\ .
\end{aligned}
\end{equation}
where we have used \eqref{FT_detT} along with the identity $\det T^{i}_a=\det(f_a^{j}T^{i}_{j})=(\det f_a^{j})(\det T^{i}_{j})=(\det T^{i}_{j})/(\det f^a_{j})$.
The second term is proportional to the trace of the undeformed CFT stress tensor. To see this, consider the saddle-point approximation, under which the undeformed CFT stress tensor satisfies
\begin{align}\label{FT_d_CFTstress}
\det e^a_i\,(T_{\rm cft})^i_a={\delta S_{\rm CFT}[e]\over \delta e^a_i}=-{1\over\mu}\epsilon^{ij}\epsilon_{ab}\left(e^b_{j}-f^b_{j}\right)\ .
\end{align}
Using the standard relation for the trace, $(T_{\rm cft})^{i}_{i}=e^a_{i}(T_{\rm cft})^{i}_a=-{c\over 24\pi}R(h^{\ast})$, where $h^\ast$ denotes the saddle-point metric, we arrive at
\begin{equation}
\begin{aligned}\label{TTbarflowEqn}
T^{i}_{i}&=2\mu\det T^{i}_{j}-\left({\det (e^{\ast})^a_i\over \det f^a_{i}}\right){c\over 24\pi}R(h^{\ast})\ ,
\end{aligned}
\end{equation}
where we have made explicit that the zweibein is evaluated at the saddle point. To leading order in~$\mu$, the anomaly term reduces to $-\frac{c}{24\pi} R(\gamma)$, and
additional corrections may arise from fluctuations around the saddle point.

To this end, using~\eqref{FT_detT}, one finds further support for the equivalence between the two descriptions: the massive gravity action reproduces the $T\bar{T}$ deformation,
\begin{align}
S_{\rm mG}[e^{\ast}, f] = \mu \int d^2x \sqrt{\gamma}  \det T^i_{j} = \mu \int d^2x \sqrt{h^{\ast}}  \det (T_{\rm cft})^i_{j}\ ,
\end{align}
where the equality follows from the identity $(\det f^a_i)\det T^i_{j} = (\det (e^{\ast})^a_i)\det (T_{\rm cft})^i_{j}$ implied by~\eqref{FT_d_stress} and~\eqref{FT_d_CFTstress}.
The second expression captures the $T\bar{T}$ deformation as the undeformed CFT evaluated on a $T\bar{T}$-deformed background geometry \cite{Conti:2018tca, Cardy:2019qao, Hirano:2024eab} -- an interpretation that, as we will see, naturally connects to the AdS/CFT correspondence.

\section{A constructive approach to the gravitational dual -- $T\Tb$ braneworld holography}
\label{sec:GD}

Assuming the AdS/CFT correspondence~\cite{Maldacena:1997re,Witten:1998qj,Gubser:1998bc}, the holographic dictionary provides a direct translation between the $T\bar{T}$-deformed CFT and its gravitational dual. The massive gravity formulation offers a concrete framework for studying holography with dynamical boundary gravity, extending the approach of~\cite{Hirano:2020nwq} to finite $T\bar{T}$ coupling. This setup may be interpreted as a realization of a Randall-Sundrum-type braneworld~\cite{Randall:1999vf, Karch:2000ct} within the AdS/CFT framework~\cite{Gubser:1999vj}, albeit with a particularly simple form of boundary gravity. In this context, the holographic dictionary maps the field theory action~\eqref{FT_action} to the following gravitational action:
\begin{align}\label{dual_massive}
S_{T\Tb\, {\rm gravity}}[g_{\alpha\beta}| e^a_i,f^a_{i}]=S_{AdS_3}[g_{\alpha\beta}|e^a_{i}]
+{1\over 2\mu}\int_{\p AdS_3} d^2x\epsilon^{ij}\epsilon_{ab}(e^a_{i}-f^a_{i})(e^b_{j}-f^b_{j})\ ,
\end{align}
where $g_{\alpha\beta}$ is the bulk metric in three dimensions, whose boundary value induces the metric $h_{ij} = \delta_{ab} e^a_i e^b_j$, with $e^a_i$ treated as dynamical.
The reference zweibein $f^a_i$ encodes the fixed, zeroth-order background geometry on the boundary.
The three-dimensional gravitational action is given by\footnote{We set the AdS radius to unity, $\ell_{\rm AdS} = 1$, so that $G = G_3/\ell_{\rm AdS}$ becomes the dimensionless Newton constant.}
\begin{align}\label{3Dgrav_action}
S_{AdS_3}[g_{\alpha\beta}|e^a_{\mu}]=-{1\over 16\pi G}\int_{AdS_3} d^3x\sqrt{g}(R+2)-{1\over 8\pi G}\int_{\del AdS_3} d^2x\sqrt{h(e)}(K(e)-1)\ ,
\end{align} 
where $K(e)$ is the extrinsic curvature constructed from the boundary zweibein $e^a_i$, and $h_{ij}(e)$ is the induced boundary metric.
 
It is important to acknowledge the work of~\cite{Caputa:2020lpa}, who were the first to apply the massive gravity description of $T\bar{T}$-deformed CFTs to holography.
In contrast to their approach, we interpret the gravitational dual action -- the LHS of~\eqref{dual_massive} -- as defined over the entire $AdS_3$ spacetime, without imposing a finite radial cutoff.
As we discuss later, the surface often interpreted as the cutoff instead emerges dynamically as a characteristic bulk surface where the deformation is neutralized, a direct consequence of the boundary gravity being dynamical.

Notably, this $T\Tb$ braneworld holography remains applicable even in the presence of matter.
The observables of the $T\bar{T}$-deformed theory can be obtained from those of the undeformed theory via
\begin{align}\label{observables_gravity_dual}
\hspace{-.3cm}
{\cal O}^{T\Tb\,{\rm gravity}}_{\rm obs}(f^a_i)
=\int {\cal D}e^a_i\, {\cal O}^{AdS}_{\rm obs}(e^a_i)\, e^{-S_{\rm mG}[e, f]}\ ,
\end{align}
where ${\cal O}^{T\bar{T}\,{\rm gravity}}_{\rm obs}$ and ${\cal O}^{\rm AdS}_{\rm obs}$ denote observables in the $T\bar{T}$-deformed gravitational dual and the undeformed AdS gravity theory, respectively. 
This extends the idea of~\cite{Hirano:2020nwq} to finite $T\bar{T}$ coupling and may be interpreted as a Gaussian ensemble of AdS gravity theories.
More specifically, the boundary correlators can be computed using the GKP-W prescription~\cite{Witten:1998qj,Gubser:1998bc}, suitably adapted to account for boundary dynamics. Translating the field-theoretic prescription developed in our previous work~\cite{HR_Correlators_mGrav} into the gravitational dual, and up to normalization, the expression takes the form
\begin{align}\label{npt_gravity_dual}
\hspace{-.3cm}
\left\langle \prod_{A=1}^N{\cal O}_{\Delta_A}(x_A)\right\rangle_{T\Tb}\!\!
=\int\! {\cal D}\alpha^i e^{-S_{\rm mG}[\alpha,f]}\prod_{A=1}^N{\delta\over \delta\phi_0(x_A\!+\!\alpha_A)}\!\int\!{\cal D}g_{\alpha\beta}{\cal D}\phi\,
e^{-S_{\rm gravity}[g_{\alpha\beta},\phi|\phi_0]}\biggr|_{\phi_0=0}\!\!\!\!
\end{align}
where the gravity action $S_{\rm gravity}[g_{\alpha\beta},\phi|\phi_0]=S_{AdS_3}[g_{\alpha\beta}]+S_{\rm matter}[\phi|\phi_0]$, and $\phi(x)$ collectively denotes bulk scalar fields with mass $m_A^2R_{AdS}^{2}=\Delta_A(\Delta_A-d)$ with $d=2$, and $\phi_0(x)$ is their boundary value, dual to the source for the primary operator of dimension $\Delta_A$.\footnote{In this setting, the three-dimensional gravity action $S_{AdS_3}[g_{\alpha\beta}]$ carries no dependence on $e^a_i = (\Phi, \alpha^i)$.}
A key point is that the boundary metric is dynamical and fluctuating, and hence the operator insertion points also fluctuate. The fields $\alpha_A^i \equiv \alpha^i(x_A)$ encode these fluctuations and appear in a specific parametrization of the boundary metric: $ds^2=e^{2\Phi}\delta_{ij}d(x^i+\alpha^i)d(x^j+\alpha^j)$. They also enter the massive gravity action as $S_{\rm mG}={1\over 4\mu}\int d^2x(\alpha^i\Box\alpha_i+\cdots)$.
We refer the reader to~\cite{HR_Correlators_mGrav} for a detailed derivation and justification of this prescription.

We emphasize that, assuming the validity of the GKP-W dictionary for the undeformed AdS/CFT correspondence and the massive gravity framework, the prescription~\eqref{npt_gravity_dual} is guaranteed to hold, thereby successfully incorporating bulk matter into our holographic dual of the $T\bar{T}$-deformed CFTs.

In parallel with the discussion in the previous section, we now turn to the holographic stress tensor in our braneworld framework. 
We work in the Fefferman-Graham gauge~\cite{Skenderis:1999nb}:\footnote{\label{foot1}In the absence of matter, the requirement that the spacetime be locally $AdS_3$ determines the form of the induced metric on a constant-radial slice and ensures that it terminates at ${\cal O}(\rho^4)$. When matter is present, however, logarithmic and higher-order corrections generally arise, and the Brown-York tensor~\eqref{holographic_stress_tensor} receives additional contributions from matter sources on the boundary~\cite{deHaro:2000vlm}. Nevertheless, these details do not affect the validity of the derivation of the $T\bar{T}$ flow equation, as is evident from the way it is obtained.}
\begin{align}\label{FG2}
ds_{AdS_3}^2={d\rho^2\over \rho^2}+{1\over\rho^2}\left(g^{(0)}_{ij}+\rho^2 g^{(2)}_{ij}+{\rho^4\over 4}(g^{(2)}g_{(0)}^{-1}g^{(2)})_{ij}\right)dx^idx^j\ ,
\end{align}
where $g^{(0)}_{ij}=h_{ij}$ denotes the boundary metric. For clarity, we express it explicitly
\begin{align}
g^{(0)}_{ij}=\delta_{ab}e^a_ie^b_j\ .
\end{align}
This expression makes manifest that the boundary metric is a dynamical field in our construction.
Within this setup, the CFT stress tensor is given by the Brown-York tensor~\cite{Balasubramanian:1999re}:
\begin{align}\label{holographic_stress_tensor}
\hspace{-.55cm}
(T_{\rm cft})_{ij}={2\over\sqrt{g_{(0)}}}{\delta S_{AdS_3}\over\delta g_{(0)}^{ij}}\biggr|_{\rho=0}\!\!
={1\over 8\pi G}\left(K_{ij}-Kh_{ij}+h_{ij}\right)={1\over 8\pi G}\!\left(\Tr\left(g_{(0)}^{-1}g^{(2)}\right)g^{(0)}_{ij}-g^{(2)}_{ij}\right),\!\!\!\!
\end{align}
with the trace condition
\begin{align}\label{trace_cond}
 \Tr\left(g_{(0)}^{-1}g^{(2)}\right)=-{1\over 2}R(g^{(0)})
\end{align}
in the absence of matter sources on the boundary~\cite{Skenderis:1999nb, deHaro:2000vlm}.
The stress tensor with mixed indices is then constructed as 
\begin{align}
(T_{\rm cft})^i_a=e^j_a(T_{\rm cft})_{ij}={1\over \det e^a_i}{\delta S_{AdS_3}\over \delta e^a_{i}}\ .
\end{align}
On the other hand, the gravitational dual of the $T\bar{T}$-deformed stress tensor takes the same form as in the field theory expression~\eqref{FT_d_stress}:
\begin{align}
\label{Grav_d_stress}
T^{i}_a={1\over \det f^a_{i}}{\delta S_{T\Tb\,{\rm gravity}}\over \delta f^a_{i}}
= -{1\over \mu\det f^a_{i}}\epsilon^{ij}\epsilon_{ab}\left(e^b_{j}-f^b_{j}\right)\ .
\end{align}
It is worth emphasizing that, in our construction, the holographic stress tensor~\eqref{Grav_d_stress} is defined at the boundary, whereas in~\cite{Kraus:2018xrn, Caputa:2020lpa}, it is instead defined on a finite radial slice at $\rho_c = \sqrt{\mu/4\pi G}$, in accordance with the cutoff AdS proposal~\cite{McGough:2016lol}.
As a result, the mechanisms by which the $T\bar{T}$ flow equation arises differ fundamentally between the two approaches: in the cutoff AdS proposal, the flow equation emerges from the Hamiltonian constraint associated with radial evolution in the absence of matter~\cite{Kraus:2018xrn, Caputa:2020lpa},, whereas in our framework, it follows directly from the boundary dynamics, without relying on bulk equations of motion. As such, the presence of matter poses no obstruction in our approach.

To make this point more precise, it is sufficient to recast the field-theoretic reasoning in gravitational terms.
At the level of the saddle-point approximation, integrating over the dynamical boundary zweibein $e^a_i$ yields the condition:
\begin{align}\label{gravity_saddle_eqn}
{\delta S_{T\Tb\,{\rm gravity}}\over \delta e^a_{i}}={\delta S_{AdS_3}\over \delta e^a_{i}}+{1\over\mu}\epsilon^{ij}\epsilon_{ab}\left(e^b_{j}-f^b_{j}\right)\biggr|_{e=e^{\ast}}=0\ .
\end{align}
Without invoking the bulk equations of motion, we can then follow the same sequence of steps from~\eqref{FT_detT} to~\eqref{TTbarflowEqn}, arriving at the flow equation
\begin{equation}
\begin{aligned}\label{TTbarflowEqn_Grav}
T^{i}_{i}&=2\mu\det T^{i}_{j}-\left({\det (e^{\ast})^a_i\over \det f^a_{i}}\right){c\over 24\pi}R(h^{\ast})\ ,
\end{aligned}
\end{equation}
where central charge $c$ is related to the dimensionless Newton constant $G$  through the standard holographic dictionary, $c=3/(2G)$, and the trace condition has been used.\footnote{In connection with Footnote~\ref{foot1}, the presence of matter sources on the boundary leads to additional contributions to the conformal anomaly~\cite{Petkou:1999fv}.}

\section{The $T\bar{T}$-deformed geometry and the dynamical emergence of a characteristic bulk surface}
\label{sec:cutoff}

To clarify how our braneworld framework connects with, and departs from, the cutoff AdS proposal~\cite{McGough:2016lol} (see also~\cite{Caputa:2020lpa} for a more closely related analysis relevant to this work), we begin by examining the influence of dynamical boundary gravity on the bulk geometry in the absence of matter. Semiclassically integrating over the boundary two-dimensional gravity, weighted by the massive gravity action, generally induces a deformation of the metric on constant-radial slices. Interestingly, however, we find that to all orders in the $T\bar{T}$ coupling, the bulk surface commonly interpreted as the cutoff arises dynamically as a distinguished surface on which the deformation is effectively neutralized.

To gain this classical geometric insight, we work within the saddle-point approximation, evaluating the path integral over the boundary zweibein in a semiclassical regime.
Specifically, our goal is to determine the boundary metric at the saddle point,
\begin{align}
g^{(0)}_{ij}=\delta_{ab}(e^{\ast})^a_i(e^{\ast})^b_j
\end{align}
by solving the saddle-point equation~\eqref{gravity_saddle_eqn}:
\begin{align}\label{gravity_saddle_eqn2}
(\det (e^{\ast})^a_i)T_{\rm cft}(e^{\ast})^i_a+{1\over\mu}\epsilon^{ij}\epsilon_{ab}\left((e^{\ast})^b_{j}-f^b_{j}\right)=0\ ,
\end{align}
where the CFT stress tensor is holographically given by~\eqref{holographic_stress_tensor}.

We now proceed to solve the saddle-point equation~\eqref{gravity_saddle_eqn2} order by order in $\mu$, expanding the zweibein as
\begin{align}
(e^{\ast})^b_{j}=f^b_{j}+\mu (e^{(1)})^b_j+\mu^2 (e^{(2)})^b_j+\mu^3 (e^{(3)})^b_j+\cdots\ .
\end{align}
We focus on the case of primary interest: the $T\bar{T}$-deformed CFT on flat space, for which the reference zweibein is given by
\begin{align}
f^a_i=\delta^a_i\ .
\end{align}
The details of the derivation are presented in Appendix~\ref{app:details}.

\subsection{Leading-order indication}
\label{sec:1st}

At first order in the expansion, the saddle-point equation takes the form~\eqref{1st_saddle}:
\begin{align}
T_{\rm cft}(\delta)^i_a+\delta^i_a(e^{(1)})^b_j\delta^j_b-\delta^i_b(e^{(1)})^b_j\delta^j_a=0\ .
\end{align}
The solution is found to be
\begin{align}\label{e(1)_solution}
(e^{(1)})^a_i=\delta^a_jT_{\rm cft}(\delta)^j_i\ .
\end{align}
To this order, the boundary metric becomes
\begin{equation}
\begin{aligned}\hspace{-.2cm}
g^{(0)}_{ij}=\delta_{ab}(e^{\ast})^a_i(e^{\ast})^b_j
=\delta_{ij}+2\mu T_{\rm cft}(\delta)_{ij}+{\cal O}(\mu^2)\ .
\end{aligned}
\end{equation}
At the same time, the holographic stress tensor~\eqref{holographic_stress_tensor} implies the following relation at leading order:
\begin{align}\label{g2T}
g^{(2)}_{ij}=-8\pi GT_{\rm cft}(\delta)_{ij}\ .
\end{align}
Substituting this into the Fefferman-Graham expansion~\eqref{FG2}, we obtain the induced metric on a constant-radial slice:
\begin{equation}
\begin{aligned}
\left. ds_{AdS_3}^2\right|_{\rho\,{\rm fixed}}={1\over\rho^2}\left(\delta_{ij}+2(\mu-4\pi G\rho^2)T_{\rm cft}(\delta)_{ij}+\cdots\right)dx^idx^j\ .
\end{aligned}
\end{equation}
Our key observation is that the radial location 
\begin{align}
\rho_c^2 = {\mu\over 4\pi G}
\end{align}
corresponds to a distinguished surface where the metric becomes flat -- that is, where the deformation is effectively neutralized. Compared to the conventions in~\cite{Caputa:2020lpa}, we have $\pi^2 \lambda_{\text{theirs}} = \mu$ and $\rho_{\text{theirs}} = \rho^2$, which reproduces the same relation satisfied by the bulk surface commonly interpreted as the cutoff.

\subsection{Second-order verification}
\label{sec:2nd}

We expect that the emergence of this characteristic surface persists to all orders in the saddle-point geometry. As a next step toward demonstrating this, we now turn to the second-order solution of the saddle-point equation.
At second order, the saddle-point equation takes the form
\begin{align}
(e^{(1)})^b_j{\delta\over\delta (e^{\ast})^b_j}(\det (e^{\ast})^a_i)(T_{\rm cft}(e^{\ast}))^i_a\biggr|_{e^{\ast}=\delta}
+\delta^i_a(e^{(2)})^b_j\delta^j_b-\delta^i_b(e^{(2)})^b_j\delta^j_a=0\ .
\end{align}
As detailed in Appendix~\ref{app:details}, the solution to this equation is found to be
\begin{align}\label{e(2)_solution}
(e^{(2)})^a_i=\delta^a_i\Tr  T^2_{\rm cft}(\delta)^2-\delta^a_k T^2_{\rm cft}(\delta)^k_i\ .
\end{align}
It follows that the boundary metric to second order is
\begin{align}
g^{(0)}_{ij}=\delta_{ij}+2\mu T_{\rm cft}(\delta)_{ij}+\mu^2\left(2\delta_{ij}\Tr T^2_{\rm cft}(\delta)-T^2_{\rm cft}(\delta)_{ij}\right)+{\cal O}(\mu^3)\ .
\end{align}
Now, making use of the holomorphicity of the CFT stress tensor in flat space -- namely $T_{zz}=T(z)$, $T_{\zb\zb}=\Tb(\zb)$, and $T_{z\zb}=0$ in the complex coordinates $(z, \zb)=(x_1+ix_2, x_1-ix_2)$ -- one can show that
\begin{align}\label{quadratic_trace}
 \delta_{ij}\Tr T^2_{\rm cft}(\delta)=2T_{\rm cft}^2(\delta)_{ij}\ .
\end{align}
Substituting this into the previous expression, we arrive at the simplified boundary metric:
\begin{align}
g^{(0)}_{ij}=\delta_{ij}+2\mu T_{\rm cft}(\delta)_{ij}+3\mu^2T_{\rm cft}^2(\delta)_{ij}+{\cal O}(\mu^3)\ .
\end{align}
To compute the induced metric on the characteristic radial slice, we first use the holographic stress tensor relation~\eqref{g2T} together with the functional derivative identity~\eqref{Tcft_derivative}.\footnote{More precisely, the holographic stress tensor is related to the metric coefficient via 
\begin{align}
g^{(2)}_{ij}=-8\pi GT_{\rm cft}(e)_{ij}-{1\over 2}R(g^{(0)})g^{(0)}_{ij}\ .
\end{align}
To justify the expansion~\eqref{g(2)1st_order}, it suffices to show that  $R(g^{(0)})={\cal O}(\mu^2)$. Since the correction to the flat-space Ricci scalar $R(\delta) = 0$ is given by
\begin{align}
\delta R =\nabla_i\left(\nabla_j\delta g^{ij}-g^{ij}\nabla_j(g_{kl}\delta g^{kl})\right)-{1\over 2}Rg_{ij}\delta g^{ij}
\end{align}
and $g^{(0)}_{ij}=\delta_{ij}+2\mu T_{\rm cft}(\delta)_{ij}+{\cal O}(\mu^2)$, the first-order $\mathcal{O}(\mu)$ correction vanishes due to the tracelessness, $\Tr T_{\rm cft}(\delta)=0$, and  conservation, $\del_jT_{\rm cft}(\delta)^{ij}=0$, of the CFT stress tensor. As a consistency check, one can also compute
\begin{align}
\Tr(g_{(0)}^{-1}g^{(2)})=g_{(0)}^{ij}g^{(2)}_{ij}=(\delta^{ij}-2\mu T_{\rm cft}(\delta)^{ij}+\cdots)(-8\pi GT_{\rm cft}(\delta)_{ij}-16\pi G\mu T_{\rm cft}(\delta)^2_{ij}+\cdots)={\cal O}(\mu^2)\ ,
\end{align}
confirming that the Ricci scalar contribution enters only at $\mathcal{O}(\mu^2)$.} 
The bulk metric coefficient $g^{(2)}(e^{\ast})_{ij}$ then expands as
\begin{align}\label{g(2)1st_order}
g^{(2)}(e^{\ast})_{ij}=g^{(2)}(\delta)_{ij}-{\mu\over 4\pi G}g^{(2)}(\delta)^2_{ij}+{\cal O}(\mu^2)\ .
\end{align}
Evaluating the full bulk metric on the slice $\rho = \rho_c$, we obtain
\begin{equation}
\begin{aligned} 
\left. ds_{AdS_3}^2\right|_{\rho=\rho_c}&={1\over\rho_c^2}\left(\delta_{ij}+3\mu^2T_{\rm cft}^2(\delta)_{ij}-4\mu^2T_{\rm cft}^2(\delta)_{ij}+\mu^2T_{\rm cft}^2(\delta)_{ij}+{\cal O}(\mu^3)\right)dx^idx^j\\
&={1\over\rho_c^2}\left(\delta_{ij}+{\cal O}(\mu^3)\right)dx^idx^j\ .
\end{aligned}
\end{equation}
Thus, we confirm that the metric on the characteristic radial slice at $\rho = \rho_c$ remains flat to second order in $\mu$.

\subsection{Third-order consistency analysis}
\label{sec:3rd}

To further substantiate our claim regarding the characteristic surface at $\rho=\rho_c$, we now perform a third-order check. At this order, the Ricci curvature $R(g^{(0)})$ -- i.e., the conformal anomaly -- begins to contribute, providing a more stringent and nontrivial test. At third order, the saddle point equation takes the form
\begin{align}
(\det (e^{\ast})^a_i)T_{\rm cft}(e^{\ast})^i_a\biggr|_{{\cal O}(\mu^2)}
+\delta^i_a\Tr\, e^{(3)}-\delta^i_b(e^{(3)})^b_j\delta^j_a=0\ .
\end{align}
where we have defined $\Tr\, e^{(3)}=(e^{(3)})^b_j\delta^j_b$. As indicated, only the ${\cal O}(\mu^2)$ term in the first expression contributes at this order.

As derived in detail in Appendix~\ref{app:details}, the solution to this equation is given by
\begin{equation}
\begin{aligned}\label{e(3)_saddle}
(e^{(3)})^a_i&=-2\delta^a_i\Tr\,T_{\rm cft}^3(\delta)+2T_{\rm cft}^3(\delta)^k_i\delta^a_k-{1\over 2}\Tr\,T_{\rm cft}^2(\delta)T_{\rm cft}(\delta)^k_i\delta^a_k\\
&-{1\over 64\pi G}\delta^a_i\partial^k\biggl(3\partial_k\Tr\, T^2_{\rm cft}(\delta)-4T_{\rm cft}(\delta)^{lm}\partial_kT_{\rm cft}(\delta)_{lm}\biggr)\ .
\end{aligned}
\end{equation}
The boundary metric can then be computed to third order as
\begin{equation}
\begin{aligned}
g^{(0)}(e^{\ast})_{ij}
&=\delta_{ij}+2\mu T_{\rm cft}(\delta)_{ij}+\mu^2\left(2\Tr\,T_{\rm cft}^2(\delta)\delta_{ij}-T_{\rm cft}^2(\delta)_{ij}\right)\\
&+\mu^3\biggl[2T_{\rm cft}^3(\delta)_{ij}+\Tr\,T_{\rm cft}^2(\delta)T_{\rm cft}(\delta)_{ij}
-4\delta_{ij}\Tr\,T_{\rm cft}^3(\delta)\\
&-{1\over 32\pi G}\delta_{ij}\partial^{k}\biggl(3\partial_{k}\Tr\, T^2_{\rm cft}(\delta)-4T_{\rm cft}(\delta)^{lm}\partial_{k}T_{\rm cft}(\delta)_{lm}\biggr)\biggr]\ .
\end{aligned}
\end{equation}
Using the complex coordinates $(z,\bar{z})$ and the (anti-)holomorphic stress tensor components, $(T, \bar{T})$, along with the identity~\eqref{quadratic_trace},  we find
\begin{align}
2T_{\rm cft}^3(\delta)_{ij}=\Tr\,T_{\rm cft}^2(\delta) \, T_{\rm cft}(\delta)_{ij}\ ,
\end{align}
which, in particular, implies that $\Tr\,T_{\rm cft}^3(\delta)=0$.
Substituting this identity simplifies the boundary metric to
\begin{equation}
\begin{aligned}\label{3rd_order_boundary_metric}
g^{(0)}(e^{\ast})_{ij}&=\delta_{ij}+2\mu T_{\rm cft}(\delta)_{ij}+3\mu^2T_{\rm cft}^2(\delta)_{ij}\\
&+\mu^3\biggl[4T_{\rm cft}^3(\delta)_{ij}
-{1\over 32\pi G}\delta_{ij}\partial^{k}\biggl(3\partial_{k}\Tr\, T^2_{\rm cft}(\delta)-4T_{\rm cft}(\delta)^{lm}\partial_{k}T_{\rm cft}(\delta)_{lm}\biggr)\biggr]\ .
\end{aligned}
\end{equation}
Meanwhile, the bulk metric coefficient $g^{(2)}(e^{\ast})_{ij}$ expands as
\begin{equation}
\begin{aligned}\label{2nd_order_bulk_metric}
g^{(2)}(e^{\ast})_{ij}&=-8\pi G\biggl[T_{\rm cft}(\delta)_{ij}+2\mu T_{\rm cft}(\delta)^2_{ij}\\
&+\mu^2 \left(3T_{\rm cft}^3(\delta)_{ij}
 -{1\over 64\pi G}\delta_{ij}\partial^{m}\biggl(3\partial_{m}\Tr\, T_{\rm cft}^2(\delta)-4T_{\rm cft}(\delta)^{kl}\partial_{m}T_{\rm cft}(\delta)_{kl}\biggr)\right)\biggr]\ .
\end{aligned}
\end{equation}
We note in passing that the final term, involving derivatives in both $g^{(0)}$ and $g^{(2)}$, originates from the Ricci curvature contribution.
Evaluating the full bulk metric on the slice $\rho = \rho_c$, we find
\begin{equation}
\begin{aligned} 
\left. ds_{AdS_3}^2\right|_{\rho=\rho_c}&={1\over\rho_c^2}\left(\delta_{ij}
+4\mu^3T_{\rm cft}^3(\delta)_{ij}-6\mu^3T_{\rm cft}^3(\delta)_{ij}+2\mu^3T_{\rm cft}^3(\delta)_{ij}+{\cal O}(\mu^3)\right)dx^idx^j\\
&={1\over\rho_c^2}\left(\delta_{ij}+{\cal O}(\mu^4)\right)dx^idx^j\ ,
\end{aligned}
\end{equation}
where the cancellation of the curvature-induced terms is more straightforward and therefore omitted for brevity.
Thus, we confirm that the induced metric on the characteristic radial slice at $\rho = \rho_c$ remains flat to third order in $\mu$.
This lends strong support for our claim.

\subsection{All-order proof}
\label{sec:all}

While there is strong supporting evidence, a complete all-order proof of the claim has not yet been established. In this section, we provide such a proof.
To begin, note that both metrics, $g^{(0)}(e^{\ast})$ and $g^{(2)}(e^{\ast})$, exhibit two distinct types of contributions, as seen in~\eqref{3rd_order_boundary_metric} and~\eqref{2nd_order_bulk_metric}: terms without derivatives and terms involving derivatives. The latter arise from the Ricci curvature.
On the surface $\rho=\rho_c$, the cancellation must take place separately for the derivative and non-derivative terms.

We begin with the non-derivative terms.
The patterns in the series expansions of~\eqref{3rd_order_boundary_metric} and~\eqref{2nd_order_bulk_metric} naturally suggest the following all-order form:
\begin{equation}
\begin{aligned}
g^{(0)}(e^{\ast})_{ij}&= \sum_{n=0}^{\infty}(n+1)\mu^n T_{\rm cft}^n(\delta)_{ij}\,+\,{\rm derivative\,\,terms}\\
&=\left(1-\mu T_{\rm cft}(\delta)\right)^{-2}_{ij} \,+\,{\rm derivative\,\,terms}\ .
\end{aligned}
\end{equation}
and
\begin{equation}
\begin{aligned}
g^{(2)}(e^{\ast})_{ij}&= -8\pi GT_{\rm cft}(\delta)_i^k\sum_{n=0}^{\infty}(n+1)\mu^n T_{\rm cft}^n(\delta)_{kj}\,+\,{\rm derivative\,\,terms}\\
&=-8\pi GT_{\rm cft}(\delta)_i^k\left(1-\mu T_{\rm cft}(\delta)\right)^{-2}_{kj} \,+\,{\rm derivative\,\,terms}\ .
\end{aligned}
\end{equation}
On the characteristic surface $\rho^2=\mu/(4\pi G)$, the combination~\eqref{FG2} appearing in the induced metric becomes
\begin{equation}
\begin{aligned}
g^{\rm ind}(e^{\ast})_{ij}&\equiv g^{(0)}(e^{\ast})_{ij}+{2\mu \over 8\pi G}g^{(2)}(e^{\ast})_{ij}+{\mu^2\over (8\pi G)^2}g^{(2)}(e^{\ast})_{ik}g^{-1}_{(0)}(e^{\ast})^{kl}g^{(2)}(e^{\ast})_{lj}\\
&\qquad =\left(\delta_i^k-2\mu T_{\rm cft}(\delta)_i^k+\mu^2T^2_{\rm cft}(\delta)^k_i\right)\left(1-\mu T_{\rm cft}(\delta)\right)^{-2}_{kj}
\,+\,{\rm derivative\,\,terms}\\
&\qquad=\delta_{ij}\,+\,{\rm derivative\,\,terms}\ .
\end{aligned}
\end{equation}
Thus, neglecting the curvature-dependent contributions for the moment, the non-derivative terms alone indeed reproduce the flat metric on the surface $\rho=\rho_c$.

To incorporate the derivative terms arising from curvature, we are guided by the trace condition $\Tr(g_{(0)}^{-1}g^{(2)})=-{1\over 2}R(g^{(0)})$. Inspection suggests the following all-order generalization:
\begin{align}
g^{(0)}(e^{\ast})_{ij}&=\left(1-{\mu\over 16\pi G}R-\mu T_{\rm cft}(\delta)\right)^{-2}_{ij}\ ,\label{all_order_g(0)}\\
g^{(2)}(e^{\ast})_{ij}&=-8\pi G\left(T_{\rm cft}(\delta)_i^k+{1\over 16\pi G}R^i_k\right)\left(1-{\mu\over 16\pi G}R-\mu T_{\rm cft}(\delta)\right)^{-2}_{kj}\ ,\label{all_order_g(2)}
\end{align}
where $R$ in the inverse-squared factors denotes the Ricci tensor $R_{ij}={1\over 2}g_{ij}R$.
These expressions indeed reproduce the results in~\eqref{3rd_order_boundary_metric} and~\eqref{2nd_order_bulk_metric}, including the curvature contribution in~\eqref{Ricci_scalar_2nd}. 
A direct check shows that the induced metric is flat:
\begin{align}
g^{\rm ind}(e^{\ast})_{ij}&=\delta_{ij}\ .
\end{align}
The all-orders zweibein is
\begin{align}\label{all_order_zweibein}
(e^{\ast})^a_i=\delta^a_k\left(\left(1-{\mu\over 16\pi G}R(g^{(0)}(e^{\ast}))-\mu T_{\rm cft}(\delta)\right)^{-1}\right)^k_i
\equiv\delta^a_k\left((1-\mu X)^{-1}\right)^k_i\ ,
\end{align}
which reproduces~\eqref{e(1)_solution}, \eqref{e(2)_solution}, and~\eqref{e(3)_saddle}.\footnote{We used $\delta^a_i\Tr T_{\rm cft}^2(\delta)=2\delta^a_k\Tr T_{\rm cft}^2(\delta)^k_i$, $2T_{\rm cft}^3(\delta)^k_i=\Tr T_{\rm cft}^2(\delta)T_{\rm cft}(\delta)^k_a$, and $\Tr T_{\rm cft}^3(\delta)=0$.} 

To establish the conjecture, it remains to verify its consistency with the saddle-point equation~\eqref{gravity_saddle_eqn2}, which can be rewritten as
\begin{align}\label{saddle_eqn_allorder}
\mu(\det (e^{\ast})^a_i)\delta^a_k\delta^b_iT_{\rm cft}(e^{\ast})^k_b+\delta^a_i\Tr\left(e^{\ast}-\delta\right)
-\left((e^{\ast})^a_{i}-\delta^a_{i}\right)=0\ ,
\end{align}
where we have defined $\Tr(e^{\ast}-\delta)=\left((e^{\ast})^b_{j}-\delta^b_{j}\right)\delta^j_b$. 
From~\eqref{holographic_stress_tensor}, the conjectured form of the CFT stress tensor is
\begin{align}
T_{\rm cft}(e^{\ast})^k_b&=(e^{\ast})^j_bT_{\rm cft}(e^{\ast})^k_j
=\left(X^k_j- \delta^k_j\Tr X\right)(1-\mu X)^j_l\delta^l_b\ ,
\end{align}
where $\Tr X={\mu\over 16\pi G}R$.
Contracting~\eqref{saddle_eqn_allorder} with $\delta^i_a$ yields the trace condition
\begin{align}\label{trace_all_order}
\Tr(e^{\ast}-\delta)=-\mu(\det (e^{\ast})^a_i)T_{\rm cft}(e^{\ast})^k_b\delta^b_k\ .
\end{align}
To check this against our conjecture, we use the identity
\begin{align}
\Tr\left(\mu X(1-\mu X)^{-1}\right)=\frac{\mu\Tr X - 2\mu^2 \det X}{\det(1-\mu X)}
\end{align} 
which follows by differentiating $\Tr\ln (1-\mu X)=\ln\det(1-\mu X)$ with respect to $\mu$, and using $\det(1-\mu X)=1-\mu\Tr X+\mu^2\det X$.
We then find
\begin{align}
T_{\rm cft}(e^{\ast})^k_b\delta^b_k=-\Tr X+2\mu\det X
\end{align}
which follows from $(\Tr X)^2-\Tr X^2=2\left({R\over 32\pi G}\right)^2-\Tr T_{\rm cft}^2(\delta)=2\det X$,
using $R^i_j={1\over 2}R\delta^i_j$ and $\det T_{\rm cft}(\delta)=-{1\over 2}\Tr T_{\rm cft}^2(\delta)$.
Noting that $(e^{\ast})^a_i-\delta^a_i=\mu X^j_k((1-\mu X)^{-1})^k_i\delta^a_j$ and $\det (e^{\ast})^a_i=1/\det (1-\mu X)$, 
we see that~\eqref{trace_all_order} is indeed satisfied in the conjectured form. 
Finally, combining these results, the saddle-point equation~\eqref{saddle_eqn_allorder} implies\footnote{The explicit inverse of $1-\mu X$ is 
\begin{align}
(1-\mu X)^{-1}={1\over \det (1-\mu X)}
\begin{pmatrix}
1-\mu\left(T_{\rm cft}(\delta)^2_2+{1\over 16\pi G}R^2_2\right) & \mu T_{\rm cft}(\delta)^1_2 \\
\mu T_{\rm cft}(\delta)^2_1 & 1-\mu\left(T_{\rm cft}(\delta)^1_1+{1\over 16\pi G}R^1_1\right)
\end{pmatrix}\ .
\end{align}}
\begin{equation}
\begin{aligned}
(e^{\ast})^a_{i}-\delta^a_{i}&=\mu\delta^a_k
\frac{X^k_i-\mu\left((X^2)^k_i-X^k_i\Tr X+2\delta^k_i\det X\right)}{\det(1-\mu X)}
=\mu\delta^a_k
\frac{X^k_i-\mu\delta^k_i\det X}{\det(1-\mu X)}\ ,
\end{aligned}
\end{equation}
where we have used the identity $(X^2)^k_i-X^k_i\Tr X=-\delta^k_i\det X$.
One can verify that this equals $\mu X(1-\mu X)^{-1}$, using $T_{\rm cft}(\delta)^1_1=-T_{\rm cft}(\delta)^2_2$ and $R^1_1=R^2_2$.
This concludes the proof.

In summary, we have demonstrated that the characteristic surface at $\rho=\rho_c$ persists to all orders in the saddle-point geometry.
The all-order expressions for the zweibein and the metrics are given in~\eqref{all_order_zweibein}, \eqref{all_order_g(0)}, and \eqref{all_order_g(2)}.
Within the $T\Tb$ braneworld holography framework, the radial ``cutoff'' surface is not imposed by hand but emerges dynamically as a distinguished bulk surface where the 
$T\Tb$ deformation is intrinsically neutralized, while the bulk spacetime continues smoothly beyond this surface to the asymptotic AdS boundary.
It is worth noting that fluctuations of the zweibein around the saddle-point configuration generate additional corrections to the bulk geometry.

\section{Derivation of the $T\bar{T}$ deformation as dynamical coordinate transformations}\label{sec:DCT}

To further strengthen the foundation of our braneworld approach to $T\bar{T}$ holography, we present a derivation that interprets the $T\bar{T}$ deformation as arising from dynamical coordinate transformations~\cite{Conti:2018tca, Cardy:2019qao}. (See also related discussions in~\cite{Hirano:2024eab}.)

First, observe that the induced metric on a generic constant radial slice takes the particularly simple form
\begin{align}
ds_{\rm ind}^2=g^{\rm ind}_{ij}(e^{\ast})dx^idx^j
=\left((1-4\pi G\rho^2 X)^2\right)_{i}^k\left(1-\mu X\right)^{-2}_{kj}dx^idx^j\ .
\end{align}
This structure suggests that the effect of the $T\bar{T}$ deformation can be reinterpreted as a coordinate transformation, $Y^i \mapsto x^i$:
\begin{align}\label{xtoYmap}
dY^i=\left((1-\mu X)^{-1}\right)^i_jdx^j\ .
\end{align}
Under this map, and using the commuting property of the symmetric matrix $M(\alpha) \equiv 1 - \alpha X$, namely $M(\alpha)M(\beta) = M(\beta)M(\alpha)$, the induced metric takes the initial Fefferman-Graham form:
\begin{align}
ds_{\rm ind}^2=(1-4\pi G\rho^2 X)^2_{ij}dY^idY^j\ ,
\end{align}
demonstrating that the $T\bar{T}$ deformation is fully absorbed into the redefined coordinates $Y^i$.

We now establish that the coordinate transformation~\eqref{xtoYmap} is precisely the dynamical coordinate transformation proposed in~\cite{Conti:2018tca}, which generalizes the infinitesimal version of~\cite{Cardy:2019qao}:
As a byproduct of our analysis in the previous section, combining~\eqref{FT_d_stress} and~\eqref{FT_d_CFTstress}, we obtain an all-order expression for the $T\bar{T}$-deformed stress tensor on $\mathbb{R}^2$ in the saddle-point approximation:
\begin{equation}
\begin{aligned}\label{TTbarT_and_cftT}
T(e^{\ast})^{i}_a&=(\det (e^{\ast})^b_{j})T_{\rm cft}(e^{\ast})^{i}_a\\
&=\left(1-\mu\Tr X+\mu^2\det X\right)^{-1}\left(X^i_j- \delta^i_j\Tr X\right)(1-\mu X)^j_k\delta^k_a
\end{aligned}
\end{equation}
where $X^i_j=T_{\rm cft}(\delta)^i_j+{1\over 16\pi G}R(e^{\ast})^i_j$. The corresponding form with spacetime indices is
\begin{align}
T(e^{\ast})^{i}_j=f^a_jT(e^{\ast})^{i}_a=\left(1-\mu\Tr X+\mu^2\det X\right)^{-1}\left(X^i_k- \delta^i_k\Tr X\right)(1-\mu X)^k_j\ ,
\end{align}
where $f^a_j=\delta^a_j$ on $\mathbb{R}^2$.
A caveat is that since the curvature $R$ is a functional of $(e^{\ast})^a_i$, obtaining a closed expression purely in terms of the flat-space CFT stress tensor $T_{\rm cft}(\delta)$ still requires solving either the saddle-point equation~\eqref{saddle_eqn_allorder} or the all-order zweibein relation~\eqref{all_order_zweibein}.
From this expression one can establish the identity
\begin{align}
\left(1-\mu T(e^{\ast})\right)^i_j={(1-\mu X)^i_j\over\det(1-\mu X)}\ ,
\end{align}
using the fact that $\left((1-\mu X)^{-1}\right)^i_j=\frac{\delta^i_k+\mu(X^i_k-\delta^i_k\Tr X)}{\det(1-\mu X)}$ for a symmetric matrix $X$.
This further implies that
\begin{align}\label{TXidentity}
\epsilon^{ik}\epsilon_{jl}\left(1-\mu T(e^{\ast})\right)^l_k={\epsilon^{ik}\epsilon_{jl}(1-\mu X)^l_k\over\det(1-\mu X)}=\left((1-\mu X)^{-1}\right)^i_j\ ,
\end{align}
since the contraction of two Levi-Civita symbols exchanges indices $1$ and $2$ and flips the sign of off-diagonal components.

Thus, the coordinate transformation~\eqref{xtoYmap} can be written equivalently as
\begin{align}\label{DCT}
dY^i=dx^i-\mu\epsilon^{ik}\epsilon_{jl}T(e^{\ast})^l_kdx^j\ .
\end{align}
This completes the holographic derivation of the dynamical coordinate transformation, extending the infinitesimal version of~\cite{Cardy:2019qao, Caputa:2020lpa} to its finite form~\cite{Conti:2018tca}.

As a final remark, we note that, using the identity~\eqref{TXidentity}, the deformation of the boundary metric can be expressed as
\begin{equation}
\begin{aligned}\label{deformed_bndy_metric_T}
g^{(0)}(e^{\ast})^i_j=\delta^i_j-2\mu\left(\delta^i_j\,\Tr\, T(e^{\ast})-T(e^{\ast})^i_j\right)+\mu^2\left(\delta^i_j\,\Tr\,T(e^{\ast})^2-T^2(e^{\ast})^i_j\right)\ .
\end{aligned}
\end{equation}
This structure resembles the ``mixed boundary conditions'' of~\cite{Guica:2019nzm}, but with an essential difference: here the stress tensor is the $T\bar{T}$-deformed one, rather than that of the undeformed CFT.

\section{Discussion}\label{sec:Discussion}

In this paper, we have proposed a constructive gravitational dual for $T\Tb$-deformed CFTs, grounded in their two-dimensional gravity description~\cite{Tolley:2019nmm}. This framework is conceptualized as a Randall-Sundrum-type braneworld~\cite{Randall:1999vf, Karch:2000ct,Gubser:1999vj}, where two-dimensional gravity localized on the $AdS_3$ boundary is dynamical.

A significant advantage of this $T\Tb$ braneworld holography is its validity and robustness even in the presence of matter, which has been a persistent limitation in alternative holographic interpretations, such as the widely assumed ``cutoff AdS proposal''~\cite{McGough:2016lol}. Our approach demonstrates how the holographic dictionary continues to hold, even with matter fields, extending previous efforts to finite $T\Tb$ coupling~\cite{Hirano:2020nwq}. We show that observables of the $T\Tb$-deformed gravitational dual can be obtained from those of the undeformed AdS gravity theory through an integration over the dynamical boundary zweibein. This provides a concrete and solvable model in which bulk geometry and matter fields emerge dynamically from boundary data under irrelevant deformations.

A central achievement of our framework is to clarify the precise link between the $T\bar{T}$ deformation and the bulk geometry, showing that the surface often viewed as a finite radial cutoff is not imposed by hand but emerges naturally at $ \rho_c^2 = \mu / (4\pi G)$ as a characteristic bulk surface where the deformation is intrinsically neutralized. Crucially, we have established an all-order proof of this dynamical emergence within the saddle-point approximation. In addition, fluctuations of the zweibein around the saddle-point configuration induce further modifications to the bulk geometry. This perspective sets our approach apart from the ``cutoff AdS proposal,'' in which the cutoff is an ad-hoc boundary that does not consistently accommodate bulk matter fields and overlooks essential features of short-distance physics below the $T\bar{T}$ scale~\cite{HR_NP_Planck}. Our dynamical realization sharpens the gravitational interpretation of the $T\bar{T}$ deformation, showing that the bulk spacetime extends smoothly beyond this characteristic surface to the asymptotic AdS boundary.

\medskip
Looking ahead, our findings open several promising avenues for future research:

\medskip\noindent
{\bf $ \bullet $ Incorporation and worked example of dynamical matter fields}:
Our framework is robust and remains valid in the presence of matter. The prescription for computing boundary correlators -- established in our earlier work~\cite{Hirano:2020nwq, HR_Correlators_mGrav}-- is already integrated into the current holographic dictionary.
The next key step is to explicitly demonstrate and apply this framework in scenarios with fully dynamical matter fields, focusing in particular on investigating the apparent coincidence between the $T\Tb$-deformed stress tensor, defined on the boundary, and the bulk Brown-York tensor in this setting. Such an explicit treatment would further reinforce and extend the applicability of the holographic dictionary.

\medskip\noindent
{\bf $\bullet$ Deformation of the characteristic surface with matter}: It would be worthwhile to examine how the dynamically emergent characteristic surface at $\rho = \rho_c$ is altered when dynamical matter fields are incorporated into the braneworld framework. In technical terms, the two-dimensional metric on constant-radial slices in~\eqref{FG2} acquires matter-induced corrections, and the trace condition~\eqref{trace_cond} is correspondingly modified in the presence of matter~\cite{deHaro:2000vlm}. These changes directly affect the saddle-point value of the zweibein derived in Section~\ref{sec:cutoff}, and thus the location and properties of the characteristic surface. Such an analysis could yield deeper insight into how matter influences the gravitational realization of the $T\bar{T}$ deformation.

\medskip\noindent
{\bf $\bullet$ Relation to other holographic proposals}: Further analysis could explore the detailed relationships and differences between our $T\Tb$ braneworld holography and other proposals in the literature, such as the ``mixed boundary proposal''~\cite{Guica:2019nzm}. At face value, our framework appears to differ substantially from that proposal, particularly due to the form of the deformed boundary metric~\eqref{deformed_bndy_metric_T} obtained by integrating out the boundary dynamical gravity. Clarifying these differences would help situate our work more precisely within the broader landscape of $T\bar{T}$ holography.

\medskip\noindent
{\bf $\bullet$ Insights into short-distance physics}: Given that the cutoff AdS proposal misses essential features of short-distance physics below the $T\Tb$ scale~\cite{HR_NP_Planck}, our constructive framework might offer new insights into these previously elusive aspects.

\medskip
We hope that our findings contribute significantly to the broader understanding of bulk reconstruction and the fundamental role of irrelevant deformations in holography.


\section*{Acknowledgments}

SH would like to thank the department of mathematics at Nagoya University for their hospitalities during his visits where part of this work was done. The work of SH is supported in part by the National Natural Science Foundation of China under Grant No.12147219.

\appendix

\section{Details of the saddle-point solution}
\label{app:details}

In this appendix, we provide a detailed derivation of the saddle-point solution to equation~\eqref{gravity_saddle_eqn2}:\footnote{Care must be taken when raising and lowering indices. For example, $\delta^i_a(e^{\ast})^b_{j}\delta^j_b \ne (e^{\ast})^i_{a}$. The correct index structure can be deduced from the identity
\begin{equation}
\begin{aligned}\label{e_einv}
\hspace{-.25cm}
\delta^i_j&=(e^{\ast})^i_{a}(e^{\ast})^a_{j}
=\delta^i_j+\mu\left((e^{(1)})^i_af^a_j+f^i_a(e^{(1)})^a_j\right)
+\mu^2\left((e^{(1)})^i_a(e^{(1)})^a_j+(e^{(2)})^i_af^a_j+f^i_a(e^{(2)})^a_j\right)+\cdots.
\end{aligned}
\end{equation}
where we have used $f^i_af^a_j=\delta^i_j$ and $f^a_if^i_b=\delta^a_b$. This implies the inverse expansions:
\begin{align}\label{e(1)inv}
(e^{(1)})^i_b=-f^i_a(e^{(1)})^a_jf^j_b\ ,\qquad (e^{(2)})^i_a=-f^i_b(e^{(2)})^b_jf^j_a-(e^{(1)})^i_b(e^{(1)})^b_jf^j_a\ ,\qquad\cdots\ .
\end{align}}
\begin{align}\label{gravity_saddle_eqn3}
(\det (e^{\ast})^a_i)T_{\rm cft}(e^{\ast})^i_a+{1\over\mu}\epsilon^{ij}\epsilon_{ab}\left((e^{\ast})^b_{j}-f^b_{j}\right)=0\ .
\end{align}
We solve this equation order by order in a perturbative expansion in $\mu$:
\begin{align}
(e^{\ast})^b_{j}=f^b_{j}+\mu (e^{(1)})^b_j+\mu^2 (e^{(2)})^b_j+\cdots\ .
\end{align}
In this paper, we carry out this expansion explicitly to the first few orders.

\subsection{First-order saddle-point solution}
\label{app:1st}

Using $\epsilon^{ij}\epsilon_{ab}=\delta^i_a\delta^j_b-\delta^i_b\delta^j_a$, the saddle-point equation~\eqref{gravity_saddle_eqn3} at first order yields
\begin{align}\label{1st_saddle}
(\det f^a_i)T_{\rm cft}(f)^i_a+\delta^i_a(e^{(1)})^b_j\delta^j_b-\delta^i_b(e^{(1)})^b_j\delta^j_a=0\ .
\end{align}
We focus on the case of primary interest: the $T\bar{T}$-deformed CFT on flat space,
\begin{align}
f^a_i=\delta^a_i\ .
\end{align}
In this case, the first-order saddle-point condition can be rewritten as
\begin{align}
T_{\rm cft}(\delta)^a_i+\delta^a_i\Tr e^{(1)}-(e^{(1)})^a_i=0\qquad\quad\mbox{with}\qquad\quad \Tr e^{(1)}\equiv (e^{(1)})^b_j\delta^j_b\ .
\end{align} 
Taking the trace of this equation then yields
\begin{align}
\Tr\, T_{\rm cft}(\delta)+\Tr e^{(1)}=0\ .
\end{align}
Since $\Tr\,e^{(1)}=-\Tr\,T_{\rm cft}(\delta)={1\over 16\pi G}R(\delta)=0$ at this order, we conclude that
\begin{align}\label{app:e(1)}
(e^{(1)})^a_i=\delta^a_jT_{\rm cft}(\delta)^j_i\ .
\end{align}

\subsection{Second-order saddle-point solution}
\label{app:2nd}

At second order in the expansion, the saddle-point equation becomes
\begin{align}\label{app:2nd_saddle_eqn}
(e^{(1)})^b_j{\delta\over\delta (e^{\ast})^b_j}(\det (e^{\ast})^a_i)T_{\rm cft}(e^{\ast})^i_a\biggr|_{e^{\ast}=\delta}
+\delta^i_a(e^{(2)})^b_j\delta^j_b-\delta^i_b(e^{(2)})^b_j\delta^j_a=0\ .
\end{align}
We can rewrite this equation as
\begin{align}
(e^{(1)})^c_j{\delta\over\delta (e^{\ast})^c_j}(\det (e^{\ast})^a_i)\delta^a_k\delta^b_iT_{\rm cft}(e^{\ast})^k_b\biggr|_{e^{\ast}=\delta}
+\delta^a_i\Tr\, e^{(2)}-(e^{(2)})^a_i=0\ .
\end{align}
where we have defined $\Tr e^{(2)} \equiv (e^{(2)})^b_j \delta^j_b$.
We now compute the first term. Using the identity $\Tr \ln e^a_i = \ln \det e^a_i$, we infer that $\delta\det (e^{\ast})^a_i/\delta (e^{\ast})^c_j=(\det (e^{\ast})^a_i)(e^{\ast})^j_c$, which gives
\begin{equation}
\begin{aligned}\hspace{-.0cm}
(e^{(1)})^c_j{\delta\over\delta (e^{\ast})^c_j}(\det (e^{\ast})^a_i)\delta^a_k\delta^b_iT_{\rm cft}(e^{\ast})^k_b\biggr|_{e^{\ast}=\delta}
&=T_{\rm cft}(\delta)^a_i \Tr e^{(1)}+\delta^a_k\delta^b_i(e^{(1)})^c_j{\delta T_{\rm cft}(e^{\ast})^k_b\over\delta (e^{\ast})^c_j}\biggr|_{e^{\ast}=\delta}\\
&=\delta^a_k\delta^b_i(e^{(1)})^c_j{\delta T_{\rm cft}(e^{\ast})_b^k\over\delta (e^{\ast})^c_j}\biggr|_{e^{\ast}=\delta}\ ,
\end{aligned}
\end{equation}
since $\Tr e^{(1)} =- \Tr T_{\rm cft}(\delta) = 0$. A useful shortcut for evaluating the variation of $T_{\rm cft}$ is to use the trace condition
\begin{align}
\Tr T_{\rm cft}(e) = e^b_kT_{\rm cft}(e)^k_b={\cal O}(\mu)\ ,
\end{align}
which implies
\begin{align}\label{Tcft_derivative}
\delta T_{\rm cft}(e)^k_b e^b_k +T_{\rm cft}(e)^k_b\delta e_k^b={\cal O}(\mu)
\qquad\Longrightarrow\qquad
\delta T_{\rm cft}(e)^k_b=-e_b^jT_{\rm cft}(e)^k_c\delta e^c_j+{\cal O}(\mu)\ .
\end{align}
Now, the saddle-point equation at second order takes the form
\begin{align}
-T^2_{\rm cft}(\delta)^a_i+\delta^a_i\Tr\, e^{(2)}-(e^{(2)})^a_i=0\ .
\end{align}
Taking the trace of this equation, we find
\begin{align}
\Tr e^{(2)}=\Tr\,T_{\rm cft}^2(\delta)\ .
\end{align}
We thus obtain the second-order zweibein:
\begin{align}\label{app:e(2)}
(e^{(2)})^a_i=\delta^a_i\Tr  T^2_{\rm cft}(\delta)-\delta^a_kT^2_{\rm cft}(\delta)^k_i\ .
\end{align}

\subsection{Third-order saddle-point solution}
\label{app:3rd}

At third order, the saddle-point equation becomes
\begin{align}\label{app:3rd_saddle_eqn}
(\det (e^{\ast})^a_i)T_{\rm cft}(e^{\ast})^i_a\biggr|_{{\cal O}(\mu^2)}
+\delta^i_a\Tr\, e^{(3)}-\delta^i_b(e^{(3)})^b_j\delta^j_a=0\ .
\end{align}
where we have defined $\Tr\, e^{(3)}=(e^{(3)})^b_j\delta^j_b$. As indicated, only the ${\cal O}(\mu^2)$ term in the first expression contributes at this order.
We can rewrite this equation as
\begin{align}\label{app:3rd_saddle_eqn2}
(\det (e^{\ast})^a_i)\delta^a_k\delta^b_iT_{\rm cft}(e^{\ast})^k_b\biggr|_{{\cal O}(\mu^2)}+\delta^a_i\Tr\, e^{(3)}-(e^{(3)})^a_i=0\ .
\end{align}
We begin by noting the expansion of the determinant:
\begin{equation}
\begin{aligned}\label{app:determinant}
\det e^a_i=1+\mu \,\underbrace{\Tr\, e^{(1)}}_{=0}+\mu^2\left(\Tr\, e^{(2)}+\det (e^{(1)})^a_i\right)+\cdots\ .
\end{aligned}
\end{equation}
To compute the ${\cal O}(\mu^2)$ contribution in the first term of the saddle-point equation~\eqref{app:3rd_saddle_eqn2}, we must expand the CFT stress tensor $T_{\rm cft}(e^{\ast})$ to second order. For this purpose, we use the holographic stress tensor relation~\eqref{holographic_stress_tensor} and the trace condition~\eqref{trace_cond}:
\begin{align}
T_{\rm cft}(e)_{ij}&={1\over 8\pi G}\!\left(\Tr\left(g_{(0)}^{-1}g^{(2)}\right)g^{(0)}_{ij}-g^{(2)}_{ij}\right)\ ,\\
\Tr (g_{(0)}^{-1}g^{(2)})&=-{1\over 2}R(g^{(0)})\ .\label{app:trace_cond}
\end{align}
Since the second-order expression for $g^{(0)}$ is already known, we now compute the second-order correction to $g^{(2)}$, using the trace condition, in order to determine the second-order expansion of the CFT stress tensor.

To proceed, we first evaluate the Ricci scalar:
\begin{equation}
\begin{aligned}
R(g)&=-{1\over\sqrt{g}}\partial_i\left(\sqrt{g}g^{ij}\partial_j\ln\sqrt{g}\right)
=-{1\over 2}\partial_i\left(g^{ij}g^{kl}\partial_j g_{kl}\right)-{1\over 4}\left(g^{mn}\partial_ig_{mn}\right)\left(g^{kl}\partial_j g_{kl}\right)\ .
\end{aligned}
\end{equation}
Our input data are the second-order expansions of the metric $g^{(0)}$ and its inverse:
\begin{align}
g^{(0)}(e^{\ast})_{ij}&=\delta_{ij}-{\mu\over 4\pi G}g^{(2)}(\delta)_{ij}+3\left({\mu\over 8\pi G}\right)^2g_{(2)}^2(\delta)_{ij}+{\cal O}(\mu^3)\ ,\\
g_{(0)}^{-1}(e^{\ast})^{ij}&=\delta^{ij}+{\mu\over 4\pi G}g_{(2)}(\delta)^{ij}+\left({\mu\over 8\pi G}\right)^2g^2_{(2)}(\delta)^{ij}+{\cal O}(\mu^3)\ .
\end{align}
A useful intermediate quantity is
\begin{equation}
\begin{aligned}\hspace{-.4cm}
(g_{(0)}^{-1})^{kl}\partial_jg^{(0)}_{kl}
=\left({\mu\over 8\pi G}\right)^2\biggl[3\partial_j\Tr g_{(2)}^2(\delta)-4g_{(2)}(\delta)^{kl}\partial_jg^{(2)}(\delta)_{kl}\biggr]+\cdots\ ,
\end{aligned}
\end{equation}
where we have used that $\Tr g^{(2)}(\delta)=0$.
It then follows that
\begin{align}\label{Ricci_scalar_2nd}
R(g)=-{1\over 2}\left({\mu\over 8\pi G}\right)^2\delta^{ij}\partial_i\biggl[3\partial_j\Tr g_{(2)}^2(\delta)-4g_{(2)}(\delta)^{kl}\partial_jg^{(2)}(\delta)_{kl}\biggr]+{\cal O}(\mu^3)\ .
\end{align}
Using the expansion
\begin{align}\label{g(2)_2nd}
g^{(2)}(e^{\ast})_{ij}=g^{(2)}(\delta)_{ij}-{\mu\over 4\pi G}g^{(2)}(\delta)^2_{ij}+\left({\mu\over 8\pi G}\right)^2 g^{(2)}_{(2)}(\delta)_{ij}+{\cal O}(\mu^3)\ ,
\end{align}
we extract the second-order correction $ g^{(2)}_{(2)}(\delta)_{ij}$ from the trace condition~\eqref{app:trace_cond} as
\begin{align}
g^{(2)}_{(2)}(\delta)_{ij}=3g_{(2)}^3(\delta)_{ij}
 +{1\over 8}\delta_{ij}\partial^{m}\biggl[3\partial_{m}\Tr g_{(2)}^2(\delta)-4g_{(2)}(\delta)^{kl}\partial_{m}g^{(2)}(\delta)_{kl}\biggr]\ ,
\end{align}
where the tensor structure of the curvature-induced second term is fixed by the identity $R_{ij}={1\over 2}g_{ij}R$ in two dimensions. 
We can now compute the CFT stress tensor to second order:
\begin{equation}
\begin{aligned}\hspace{-.4cm}
T_{\rm cft}(e^{\ast})^i_{j}&\!=\!-{1\over 8\pi G}\biggl[g^{(2)}(\delta)^i_j
- {1\over 8}\!\left({\mu\over 8\pi G}\right)^2
\delta^i_j\partial^{m}\!\biggl(3\partial_{m}\!\!\Tr g_{(2)}^2(\delta)-4g_{(2)}(\delta)^{kl}\partial_{m}g^{(2)}(\delta)_{kl}\biggr)\biggr].\!\!\!\!\!\!\!\!\!
\end{aligned}
\end{equation}
As a consistency check, we verify that the trace satisfies
\begin{align}
\Tr\, T_{\rm cft}(e^{\ast})=-{c\over 24\pi}R(g^{(0)})\qquad\mbox{with}\qquad c={3\over 2G}\ .
\end{align}
The CFT stress tensor with mixed indices is then given, to second order, by
\begin{equation}
\begin{aligned}\label{CFTstress_mixed_3rd}
T_{\rm cft}(e^{\ast})^i_a&=(e^{\ast})^k_aT_{\rm cft}(e^{\ast})^i_{k}\\
&=T_{\rm cft}(\delta)^i_a-\mu T^2_{\rm cft}(\delta)^i_j\delta^j_a
+\mu^2\biggl[\left(2T^3_{\rm cft}(\delta)^i_j\delta^j_a-\Tr\, T^2_{\rm cft}(\delta)T_{\rm cft}(\delta)^i_a\right)\\
&+{1\over 64\pi G}\delta^i_a\partial^k\biggl(3\partial_k\Tr\, T^2_{\rm cft}(\delta)-4T_{\rm cft}(\delta)^{lm}\partial_kT_{\rm cft}(\delta)_{lm}\biggr)\biggr]
\end{aligned}
\end{equation}
where we have used the holographic stress tensor relation $T_{\rm cft}(\delta)^i_j=-{1\over 8\pi G}g^{(2)}(\delta)^i_j$ evaluated on flat space.

Combining the determinant expansion~\eqref{app:determinant} and the third-order expression for the CFT stress tensor~\eqref{CFTstress_mixed_3rd}, together with the first- and second-order zweibeins~\eqref{app:e(1)} and~\eqref{app:e(2)} as inputs, we compute the first term in the saddle-point equation~\eqref{app:3rd_saddle_eqn2}:
\begin{equation}
\begin{aligned}\hspace{-.4cm}
(\det (e^{\ast})^a_i)T_{\rm cft}(e^{\ast})^i_a\biggr|_{{\cal O}(\mu^2)}
&=-{1\over 2}\Tr T^2_{\rm cft}(\delta)T_{\rm cft}(\delta)^i_a+2T^3_{\rm cft}(\delta)^i_j\delta^j_a\\
&+{1\over 64\pi G}\delta^i_a\partial^k\biggl(3\partial_k\Tr\, T^2_{\rm cft}(\delta)-4T_{\rm cft}(\delta)^{lm}\partial_kT_{\rm cft}(\delta)_{lm}\biggr)
\end{aligned}
\end{equation}
where we have used $\Tr\, e^{(2)}=\Tr T_{\rm cft}^2(\delta)$ and $\det T_{\rm cft}(\delta)=-{1\over 2}\Tr T^2_{\rm cft}(\delta)$.
Taking the trace of the saddle-point equation~\eqref{app:3rd_saddle_eqn2}, we find
\begin{equation}
\begin{aligned}\label{e(3)trace}
\Tr\,e^{(3)}&=-2\Tr\,T_{\rm cft}^3(\delta)
-{1\over 32\pi G}\partial^i\biggl(3\partial_i\Tr\, T^2_{\rm cft}(\delta)-4T_{\rm cft}(\delta)^{lm}\partial_iT_{\rm cft}(\delta)_{lm}\biggr)\ .
\end{aligned}
\end{equation}
We thus obtain the third-order zweibein:
\begin{equation}
\begin{aligned}
(e^{(3)})^a_i&=-2\delta^a_i\Tr\,T_{\rm cft}^3(\delta)+2T_{\rm cft}^3(\delta)^k_i\delta^a_k-{1\over 2}\Tr\,T_{\rm cft}^2(\delta)T_{\rm cft}(\delta)^k_i\delta^a_k\\
&-{1\over 64\pi G}\delta^a_i\partial^k\biggl(3\partial_k\Tr\, T^2_{\rm cft}(\delta)-4T_{\rm cft}(\delta)^{lm}\partial_kT_{\rm cft}(\delta)_{lm}\biggr)\ .
\end{aligned}
\end{equation}
As a consistency check, we can verify that contracting this equation with $\delta^i_a$ yields the trace~\eqref{e(3)trace}. 

Finally, the deformed boundary metric to third order is computed as
\begin{equation}
\begin{aligned}
g^{(0)}(e^{\ast})_{ij}
&=\cdots+\mu^3\biggl(\delta_{ab}\left((e^{(1)})^a_i(e^{(2)})^b_j+(e^{(2)})^a_i(e^{(1)})^b_j\right)+(e^{(3)})^a_i\delta_{aj}+e^{(3)})^a_j\delta_{ai}\biggr)\\
&=\delta_{ij}+2\mu T_{\rm cft}(\delta)_{ij}+\mu^2\left(2\Tr\,T_{\rm cft}^2(\delta)\delta_{ij}-T_{\rm cft}^2(\delta)_{ij}\right)\\
&+\mu^3\biggl[2T_{\rm cft}^3(\delta)_{ij}+\Tr\,T_{\rm cft}^2(\delta)T_{\rm cft}(\delta)_{ij}
-4\delta_{ij}\Tr\,T_{\rm cft}^3(\delta)\\
&-{1\over 32\pi G}\delta_{ij}\partial^k\biggl(3\partial_{k}\Tr\, T^2_{\rm cft}(\delta)-4T_{\rm cft}(\delta)^{lm}\partial_{k}T_{\rm cft}(\delta)_{lm}\biggr)\biggr]\ .
\end{aligned}
\end{equation}
Meanwhile, the third-order expansion of the bulk metric coefficient $g^{(2)}(e^{\ast})_{ij}$ in~\eqref{g(2)_2nd} takes the form:
\begin{equation}
\begin{aligned}\hspace{-.3cm}
g^{(2)}(e^{\ast})_{ij}&=-8\pi G\biggl[T_{\rm cft}(\delta)_{ij}+2\mu T_{\rm cft}(\delta)^2_{ij}\\
&+\mu^2 \left(3T_{\rm cft}^3(\delta)_{ij}
 -{1\over 64\pi G}\delta_{ij}\partial^{m}\biggl(3\partial_{m}\Tr\, T_{\rm cft}^2(\delta)-4T_{\rm cft}(\delta)^{kl}\partial_{m}T_{\rm cft}(\delta)_{kl}\biggr)\right)\biggr]\ .\!\!
\end{aligned}
\end{equation}


\end{document}